\documentclass{ws-rv9x6}

\usepackage{subfigure}
\usepackage[square]{ws-rv-van}
\usepackage[usenames]{color}

\makeindex

\newcommand{\vt}[1]{{\mathbf #1}} % vector notation
\newcommand{\x}{x_\alpha(t)} % FBM of index alpha

\newcommand{\la}{\langle}
\newcommand{\ra}{\rangle}

\begin{document}

\chapter[First passage processes of FBM and diffusion-limited reaction phenomena]{First passage behaviour of multi-dimensional fractional Brownian motion and application to reaction phenomena}

\author[J.-H. Jeon, A. V. Chechkin, \& R. Metzler]{Jae-Hyung Jeon$^1$,
A. V. Chechkin$^{2,3}$, and Ralf Metzler$^{4,1}$}

\address{$^1$Department of Physics, Tampere University of Technology\\
Tampere 33101, Finland\\
$^2$ Akhiezer Institute for Theoretical Physics NSC KIPT \\
Akademicheskaya Str.1, 61108 Kharkov, Ukraine\\
$^3$ Max-Planck Institute for the Physics of Complex Systems,
N{\"o}thnitzer Stra{\ss}e 38, 01187 Dresden, Germany\\
$^4$ Institute for Physics \& Astronomy, University of Potsdam\\
14476 Potsdam-Golm, Germany\\
E-mail: jae-hyung.jeon@tut.fi}

\begin{abstract}
Fractional Brownian motion is a generalised Gaussian diffusive process that is
found to describe numerous stochastic
phenomena in physics and biology. Here we introduce a multi-dimensional
fractional Brownian motion (FBM) defined as a superposition of
conventional FBM for each coordinate in analogy to multi-dimensional
Brownian motion, and study its first passage properties. Starting from the
well-established first passage time statistics of one-dimensional FBM and the
associated approximation schemes, we explore the first passage
time behaviour of multi-dimensional FBM and compare these results with
simulations. The asymptotic kinetic behaviour of diffusion-limited reactions
of reactant particles performing FBM in a one- and multi-dimensional space is
studied based on the corresponding first passage time statistics.
\end{abstract}

\section{Introduction}

Theories of anomalous diffusion and active transport of physical particles
are becoming increasingly important concepts and quantitative tools in
many branches of the physical sciences, especially biology-related fields,
as deviations from normal free and directed Brownian motion have been
ubiquitously observed in many systems, in the wake of significant
advances of experimental techniques such as single particle tracking
\cite{havlin,bouchaud,ralfreview1,ralfreview2,xie,newby}. Anomalous dynamic
phenomena are manifested through a departure from the linear time dependence
of the mean squared displacement of Brownian motion, $\langle\mathbf{r}^2(t)
\rangle\simeq t$ \cite{havlin,bouchaud,ralfreview1}. In many cases,
data follow the scaling law
\begin{equation}
\la\mathbf{r}^2(t)\ra\propto t^\alpha,~\hbox{with}~\alpha\neq1.\label{anomsd}
\end{equation}
Here the anomalous diffusion exponent $\alpha$ is in the range $0<\alpha<1$
for subdiffusion and $1<\alpha<2$ for superdiffusion. $\alpha=2$ denotes
ballistic motion, while $\alpha>2$ is often
referred to as hyperdiffusion.
The generalised diffusion law \eqref{anomsd} is nonuniversal in the
sense that it may be based on multiple physical mechanisms and equally
well be described by several, physically different, prominent diffusion
models. More specifically, fractional Brownian motion (FBM) gives rise to
anomalous diffusion processes \eqref{anomsd} in which the displacement
autocorrelation function (DAF) has a power-law decay with a positive or negative
sign at long times. In one spatial coordinate $x$, this implies
$\mathrm{DAF}\sim\alpha(\alpha-1)t^{\alpha-2}$. Meanwhile, continuous time random walks (CTRWs)
describe processes in which the diffusion is hindered by multiple trapping
events whose duration is distributed according to $\psi(\tau)\propto\tau^{-1-
\alpha}$ with $0<\alpha<1$, such that
the mean waiting time $\la\tau\ra$ diverges \cite{ralfreview1,montroll,scher}.
Superdiffusive CTRWs are either based on the spatiotemporally coupled L{\'e}vy
walk scheme with finite moments of all orders, such that the MSD (\ref{anomsd})
is sub-ballistic with $1<\alpha<2$. Alternatively, they are described by L\'evy
flights
corresponding to CTRW processes with a broad jump length distribution of the
power-law form $\lambda(x)\propto|x|^{-1-\mu}$ with $0<\mu<2$,
leading to a diverging variance $\la x^2\ra$.\footnote{In this case,
the conventional mean squared displacement \eqref{anomsd} is replaced by
rescaled fractional order moments
$\la |x(t)|^\delta\ra^{2/\delta}\propto t^{2/\mu}$, with $0<\delta<\mu$
\cite{ralfreview1,ralfreview2,fogedby,ralfchapter}.} Within the model of
diffusion on fractals, the subdiffusive motion of the form \eqref{anomsd}
is induced by the fractal geometry of the space in which the diffusion
process takes place \cite{havlin}.

While all of the aforementioned anomalous diffusion models share the scaling
law \eqref{anomsd}, they differ in their stochastic properties and thus exhibit
distinct diffusion-controlled dynamics. The first passage behaviour is an
important dynamic measure that pinpoints stochastic subtleties of the underlying
anomalous diffusive process \cite{ralfreview2,redner}. First passage theory is
a powerful
stochastic concept that can be applied to attaining a quantitative description
of numerous dynamic problems including diffusion-limited reaction kinetics
\cite{wf1,wf2,schulten}, polymer cyclisation \cite{sokolov2,guerin,likthman},
target search \cite{lomholt,condamin,condamin2}, barrier crossing
\cite{goychuk3,oleksii}, and polymer translocation \cite{vilgis}, among many
others. In this chapter we study the first passage properties of FBM-type
anomalous diffusions with applications to diffusion-limited reaction kinetics.

First passage statistics are systematically obtained using standard methods for
Markovian processes \cite{redner}. However, their analytic derivation becomes
challenging for anomalous diffusive processes. In some cases, though, full
analytical solutions are available. This is true for subdiffusive CTRW
processes \cite{condamin2,metzlerctrw,rangarajan,ding2,condamin3}, for which
the first passage behaviour is obtained by standard
techniques such as the separation of variables or the images method.
Alternatively, it can be directly derived from the corresponding Brownian
boundary value problem by subordination \cite{ralfreview2}. For L\'evy flights
some results have been derived such as the Sparre Andersen $3/2$ scaling of
the first passage density and the slower decay of the probability density
of first arrival \cite{aleksei,koren}. For diffusion in complex environments,
general statements on the mean first passage time are available \cite{condamin}.
In the case of FBM, however, analytic approaches are notoriously
limited, the best known analytical result being Molchan's asymptotic scaling
result of the first passage behaviour of FBM in a semi-infinite, one-dimensional
interval with an absorbing boundary \cite{ding1,molchan}. Except for this
case, generally the first passage behaviour of FBM is quite poorly understood
and mostly limited to the one-dimensional case \cite{vilgis,kantor,rosso,gleb}.
Based on the fact that many dynamic processes in nature, including intracellular
diffusive processes (see below), are governed by FBM, its first passage
behaviour in higher dimensions is of particular interest and relevance. This
is especially so as the sample path of FBM has fractal dimension $2/\alpha$:
FBM thus explores the space both compactly (recurrently) and non-compactly
(transiently) for different anomalous diffusion exponents $\alpha$ when the
dimensionality of the embedding space is greater than unity.

In this chapter we collect recent theoretical results for the first passage
behaviour of FBM in one and higher spatial dimensions and discuss their
implications to diffusion-limited reaction phenomena. Our study focuses on
the anomalous reaction kinetics resulting from FBM-type anomalous diffusion
of reactant particles. Anomalous reaction problems based on other models such
as CTRW or diffusion on fractals can be found, for instance, in
Refs.~\cite{katja,turner,havlin2,yuste0,yuste,seki,sokolovreaction,froemberg,henry}.
In Sec.~\ref{sec_fbm}, we start from an overview of some basic facts on FBM 
and the related motion governed by the fractional Langevin equation, present its generalisation
to $d$-dimensional FBM, and comment on the fractal dimension of FBM. In
Sec.~\ref{sec_fpt} we define the first passage process with mathematical rigour
and then provide exact results and approximations of first passage of 
one-dimensional FBM, along with a comparison to simulations results. In the
subsequent Section, we consider the first passage problem of FBM in
$d$-dimensional space, seeking the first passage time statistics based on the
results of the one-dimensional case and the approximation theory presented
in Sec.~\ref{sec_fpt}. Consequences for diffusion-limited reaction kinetics of
FBM reactants in one and higher dimensional space are also discussed. Finally,
our concluding remarks are presented in Sec.~\ref{conclusion}.

\section{Fractional Brownian motion}\label{sec_fbm}

One-dimensional FBM $x_\alpha(t)$ with the anomalous diffusion exponent
$\alpha$ ($0<\alpha<2$), was originally studied by Kolmogorov \cite{kolmogorov}
and Yaglom \cite{yaglombook}, and later became famous through
the work of Mandelbrot and van Ness \cite{ness}. FBM may be viewed as a natural
extension of normal Brownian motion (BM). Thus, FBM is a Gaussian process
satisfying the criteria that $x_\alpha(0)=0$ and for all $t$ and $\Delta
t$ the increment $x_\alpha(t+\Delta t)-x(t)$ is stationary and given by
the normal distribution of zero mean and variance $2K_\alpha(\Delta
t)^\alpha$ so that its probability density function satisfies \cite{yaglombook}
\begin{equation}
G(x_\alpha(t+\Delta t)|x_\alpha(t))=\frac{1}{\sqrt{4\pi K_\alpha(\Delta t)^
\alpha}}\exp\left(-\frac{[x_\alpha(t+\Delta t)-x_\alpha(t)]^2}{4K_\alpha(\Delta
t)^\alpha}\right).
\label{fbmdist}
\end{equation} 
Here $K_\alpha$ is the generalised diffusivity of physical dimension $\mathrm{
cm}^2/\mathrm{sec}^\alpha$. By this definition FBM satisfies the expected properties of a generalised Brownian motion, i.e., 
\begin{equation}
\langle x^2_\alpha(t)\rangle=2K_\alpha t^\alpha
\end{equation}
as well as
\begin{equation}
\langle [x_\alpha(t+\Delta t)-x_\alpha(t)]^2\rangle=2K_\alpha(\Delta t)^\alpha.
\end{equation}
It is immediately inferred from these relations that the covariance of FBM is uniquely given by \cite{falconer,sokolov}
\begin{equation}
\langle x_\alpha(t_1)x_\alpha(t_2)\rangle=K_\alpha(|t_1|^\alpha+|t_2|^\alpha-|t_1-t_2|^\alpha).
\end{equation}
Thus, except for the special case $\alpha=1$ of ordinary BM, the increments of
FBM are correlated. FBM was shown to
quantitatively describe numerous anomalous diffusive phenomena found in
nature, for instance, annual river discharges \cite{hurst}, diffusion of a
tracer particle in a single file \cite{harris,lebowitz,wei,lizana} or in a
viscoelastic environment \cite{goychuk,goychuk2},
conformational dynamics of proteins \cite{kou}, diffusion of macromolecules in crowded or intracellular
environments \cite{vicent,jeon2,weber,weiss,natascha}, or the lateral diffusion
of phospholipid molecules in lipid bilayers \cite{kneller,jeon3}.

Mandelbrot and van Ness \cite{ness} presented an explicit form of the above FBM process via \emph{fractional integration} of a white and Gaussian noise: 
\begin{eqnarray}
\x&=&\frac{1}{\Gamma(\frac{\alpha+1}{2})}\left(\int_0^t d\tau(t-\tau)^{(\alpha-1)/2}\xi(\tau) \right.\nonumber \\ &+&\left. \int_{-\infty}^0 d\tau\Big[(t-\tau)^{(\alpha-1)/2}-(-\tau)^{(\alpha-1)/2}\Big]\xi(\tau)\right)
\label{fbmness}
\end{eqnarray}
where $\Gamma(z)$ is the Gamma function and $\xi(\tau)$ represents white Gaussian
noise with zero mean $\la\xi(t)\ra=0$ and $\delta$-correlation $\la\xi(t)\xi(t')
\ra=\sqrt{2K_1}\delta(t-t')$. Expression \eqref{fbmness} leads to FBM with the
generalised diffusivity $K_\alpha=K_1[\Gamma((\alpha+1)/2)]^{-2}\left\{\int_{-
\infty}^0[(1-\tau)^{(\alpha-1)/2}-(-\tau)^{(\alpha-1)/2}]^2d\tau+1/\alpha\right\}
$ \cite{ness}. In the special case $\alpha=1$, the integral representation of
FBM \eqref{fbmness} reduces to the familiar expression of BM, $x_1(t)=\int_0^t
d\tau\xi(\tau)$. 

\begin{figure}[tb]
\centering
\psfig{file=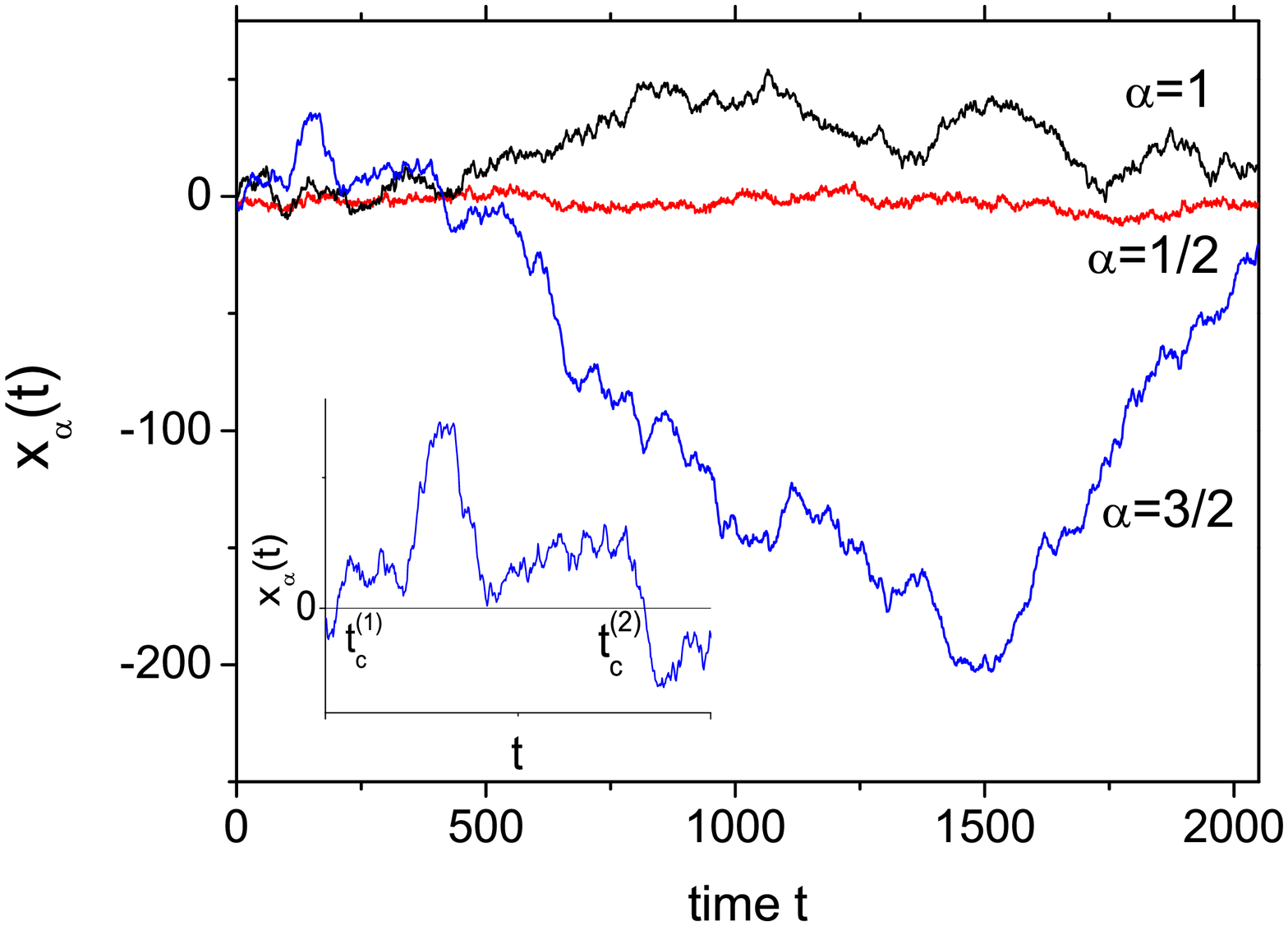,width=5.7cm}
\hspace*{-5pt}
\psfig{file=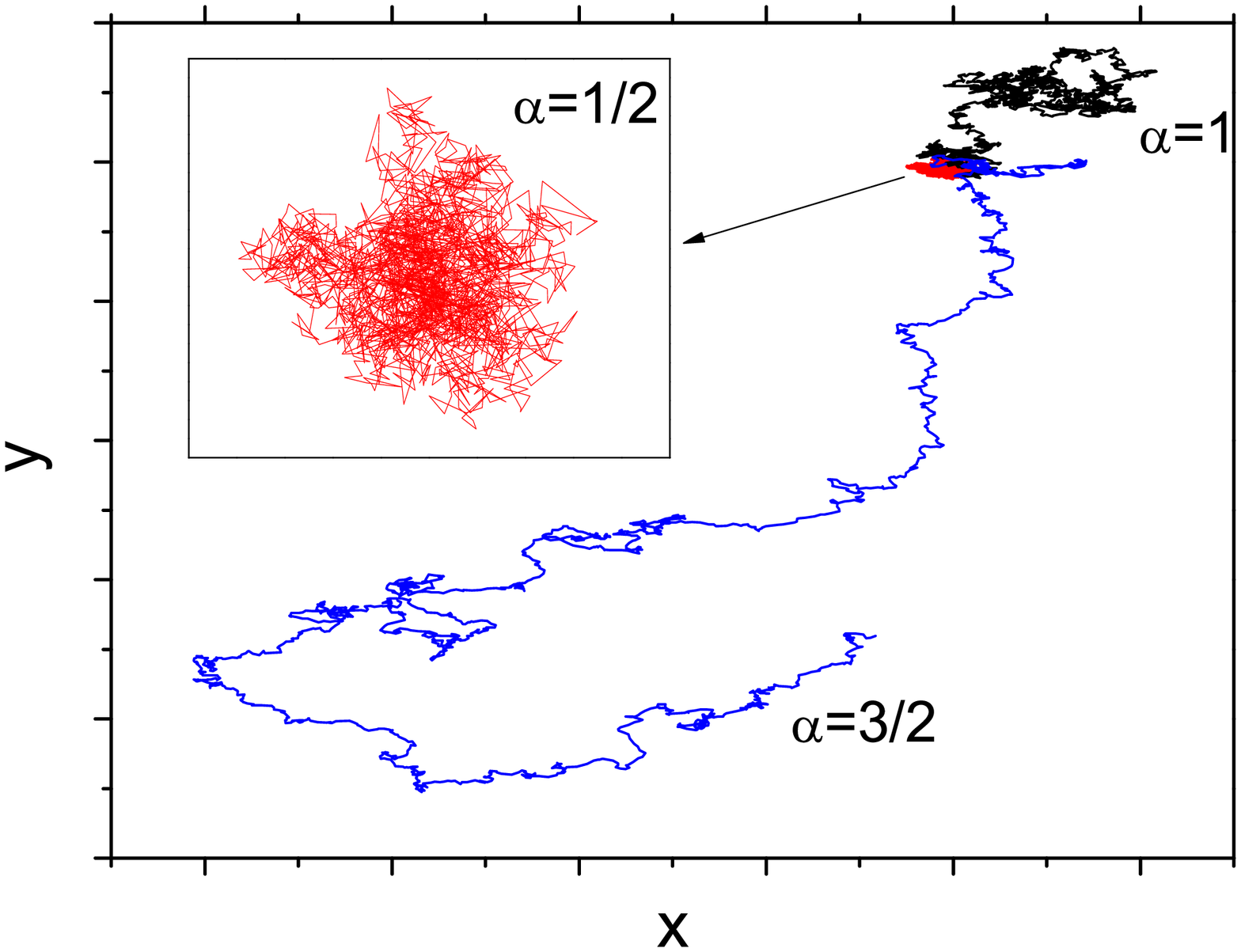,width=5.3cm}
\caption{(Left) Sample FBM trajectories $\x$ for three different cases:
subdiffusion ($\alpha=1/2$), normal diffusion ($\alpha=1$), and superdiffusion
($\alpha=3/2$). Inset: Zero-crossing times $t^{(i)}_c(>0)$ are defined as
the intersects of $\x$ and the horizontal line of $y=0$ for a given FBM
trajectory. (Right) Two-dimensional FBM trajectories $\mathbf{x}_\alpha(t)$
with the anomalous diffusion exponent $\alpha=1/2$ (red or light grey), 1
(black), 3/2 (blue or dark grey). Inset: Zoom-in motion of the subdiffusive
walk of $\alpha=1/2$.}
\label{fbm1d}
\end{figure}

Alternatively, FBM can be expressed in terms of its incremental sequence, i.e., fractional Gaussian noise (FGN) $\xi_\alpha(t)$, such that
\begin{equation}
x_\alpha(t)=\int_0^t dt'\xi_\alpha(t').\label{fbmdef2}
\end{equation}      
Here FGN $\xi_\alpha(t)$ with the anomalous diffusion exponent $0<\alpha<2$ is a generalised Gaussian process of zero mean $\langle \xi_\alpha(t)\rangle=0$ and satisfies the autocorrelation of the form \cite{ness,jeon1,qian}
\begin{eqnarray}
\label{autocorfgn}
\langle \xi_\alpha(t_1)\xi_\alpha(t_2)\rangle=\left\{\begin{array}{ll} \alpha K_\alpha(\alpha-1)|t_1-t_2|^{\alpha-2},& \alpha\neq1,~|t_1-t_2|\rightarrow\infty \\
2K_\alpha \delta(t_1-t_2), & \alpha=1\end{array}\right..
\end{eqnarray}
Note that FGN with anomalous diffusion exponent $\alpha\neq1$ has a power-law
decay of the correlations with exponent $\alpha-2$ in the long-time limit. The
prefactor tells that FGN is negatively (positively) correlated for $0<\alpha<1$
($1<\alpha<2$) and, in turn, gives rise to subdiffusive (superdiffusive)
FBM. For $0<\alpha<1$ FGN is often called antipersistent, and persistent
in the case $1<\alpha<2$. According to Eq.~(\ref{autocorfgn}) the limiting
case $\alpha=2$ of ballistic motion is fully persistent. At $\alpha=1$
FGN becomes ordinary white and Gaussian noise. Figure \ref{fbm1d}(Left)
illustrates sample paths of FBM $x_\alpha(t)$ obtained from FGN for cases of
subdiffusion ($\alpha=1/2$), normal diffusion ($\alpha=1$), and superdiffusion
($\alpha=3/2$). The significantly enhanced exploration of space for growing
$\alpha$ is distinct.

We note that in our simulations we generate FGN using the Hosking method
\cite{hosking}. Alternative methods are described in
Refs.~\cite{langowski,chegon}.

%%%%%%%%%%%%%%%%%%%%% FIGURE %%%%%%%%%%%%%%%%%%%%%%%%%%%%%%%%%%
%\begin{verbatim}
%%%%%%%%%%%%%%%%%%%%%%%%%%%%%%%%%%%%%%%%%%%%%%%%%%%%%%%%%%%%%%%%%

\subsection{Relation to fractional Langevin equation driven by FGN}

FBM is a nonequilibrium stochastic process driven by FGN in the sense that it
is not subject to the fluctuation-dissipation theorem
[see \eref{fbmdef2}]. The thermal motion $y_\alpha(t)$ of a particle of mass
$m$ driven by FGN is described by the fractional Langevin equation (FLE)
with anomalous diffusion exponent $0<\alpha<1$
\cite{jeon1,zwanzig,gemant,luts,goychuk,goychuk2}
\begin{equation}
m\frac{d^2y_\alpha(t)}{dt^2}=-\gamma \int_0^t d\tau (t-\tau)^{\alpha'-2}\frac{dy_\alpha}{d\tau}+\left(\frac{k_B\mathcal{T}\gamma}{\alpha'(\alpha'-1)K_{\alpha'}}\right)^{1/2}\xi_{\alpha'}(t).
\label{fle}
\end{equation} 
Here $\gamma(>0)$ is the generalised friction coefficient, $k_B$ the
Boltzmann constant, and $\mathcal{T}$ the absolute temperature. \Eref{fle}
is a special form of the generalised Langevin equation with power-law noise
correlations. As the integral in Eq.~(\ref{fle})
defines the fractional Caputo operator $\partial
^{2-\alpha}/\partial t^{2-\alpha}$, we can rephrase Eq.~(\ref{fle}) in the
fractional form
\begin{equation}
m\frac{d^2y_\alpha(t)}{dt^2}=-\gamma\frac{\partial^{2-\alpha}}{\partial t^{2-
\alpha}}y_{\alpha}(t)+\left(\frac{k_B\mathcal{T}\gamma}{\alpha'(\alpha'-1)K_{
\alpha'}}\right)^{1/2}\xi_{\alpha'}(t),
\end{equation}
hence the name FLE.
The above FLE is constructed with the persistent FGN  $\xi_{\alpha'}(t)$
of $1<\alpha'<2$ and obeys the fluctuation-dissipation theorem
$\eta^2\la \xi_{\alpha'}(t)\xi_{\alpha'}(t')\ra=\gamma
k_B\mathcal{T} |t-t'|^{\alpha'-2}$ where
$\eta\equiv(k_B\mathcal{T}/[\alpha'(\alpha'-1)K_{\alpha'}])^{1/2}$.\footnote{
FLE motion $y_\alpha(t)$ with anomalous diffusion exponent
$1<\alpha<2$ leading to superdiffusion in the long-time limit and satisfying
the fluctuation-dissipation theorem can be defined with the anti-persistent FGN
in a similar way, see Refs.~\cite{siegle1, siegle2}.} Using the generalised
Mittag-Leffler function $E_{a,b}(z)=\sum_{n=0}^{\infty}z^n/\Gamma(a n+b)$,
one finds that the variance of the above FLE is \cite{luts,pottier}
\begin{equation}
\la y_\alpha^2(t)\ra=2\frac{k_B\mathcal{T}}{m} t^2 E_{\alpha',3}\left[-\frac{
\gamma\Gamma(\alpha'-1)}{m}t^{\alpha'}\right]\label{flevariance}
\end{equation}  
with the assumption of thermal initial velocity $\la \dot{y}_\alpha^2(0)\ra=k_B\mathcal{T}/m$. \Eref{flevariance} states that FLE motion shows a crossover from the short-time ballistic motion $\langle y_\alpha^2(t)\rangle\sim t^2$ to the long-time subdiffusion of the exponent $0<\alpha<1$
\begin{eqnarray}
\langle y_\alpha^2(t)\rangle\sim t^{\alpha}
\end{eqnarray} 
in the overdamped limit ($t\rightarrow\infty$). Note that the subdiffusion exponent $\alpha$ of the FLE process relates with the anomalous diffusion exponent $\alpha'$ of the driving FGN in \eref{autocorfgn} through 
\begin{equation}
\alpha=2-\alpha'.\label{exponent}
\end{equation}
It turns out that the overdamped FLE process $y_\alpha(t)$ fulfils all the criteria of FBM, thus behaving as a subdiffusive FBM $\x$ of the same anomalous diffusion exponent $\alpha$ \cite{deng,taloni}.     

FLE motion was simulated via the method discussed in Ref.~\cite{jeon1}. Thus,
the stochastic Volterra integral equation of $dy(t)/dt$ derived from
Eq.~(\ref{fle}) is numerically solved, while the FGN is created with the
Hosking method. Alternative methods to generate FLE motion are found in
Refs.~\cite{tobias,goychuk,goychuk2}.

\subsection{Multi-dimensional fractional Brownian motion}

One may introduce FBM in $d$-dimensional space $\mathbf{x}_\alpha(t)$ as a superposition of independent FBM $x_\alpha(t)$ for each Cartesian coordinate \cite{yaglom3,unterberger,qian2,jeon1}, namely, 
\begin{equation}
\mathbf{x}_\alpha(t)=\sum_{i=1}^d\int_0^t dt'\xi_\alpha^{(i)}(t')\hat{x}_i
\end{equation}
where $\hat{x}_i$ is the $i$th component in the Cartesian coordinate and $\xi_
\alpha^{(i)}(t)$ is FGN of zero mean $\langle \xi_\alpha^{(i)}(t)\rangle=0$ and covariance $\langle \xi_\alpha^{(i)}(t_1)\xi_\alpha^{(j)}(t_2)\rangle=\langle \xi_\alpha(t_1)\xi_\alpha(t_2)\rangle \delta_{ij}$. From this definition, $d$-dimensional FBM $\mathbf{x}_\alpha(t)$ is shown to satisfy the three criteria \cite{jeon1}: 
\begin{subequations}
\begin{eqnarray}
\langle\mathbf{x}_\alpha(t)\rangle&=&0,\\
\langle\mathbf{x}_\alpha^2(t)\rangle&=&2dK_\alpha t^\alpha,\label{msd2}\\
\langle\mathbf{x}_\alpha(t_1)\cdot\mathbf{x}_\alpha(t_2)\rangle&=&dK_\alpha(|t_1|^\alpha+|t_2|^\alpha-|t_1-t_2|^\alpha).
\end{eqnarray}
\end{subequations}
Note that the absolute magnitude of $\mathbf{x}_\alpha(t)$, $|\mathbf{x}_\alpha(t)|$, is not a one-dimensional FBM. In analogy to this, $d$-dimensional FLE motion can be also defined via $\vt{y}_\alpha(t)=\sum_{i=1}^d y_\alpha^{(i)}(t)\hat{x}_i$ where $y_\alpha^{(i)}(t)$ is the FLE process \eqref{fle} with FGN $\xi_{\alpha'}^{(i)}(t)$ and $1<\alpha'<2$.

\subsection{Fractal properties}

FBM is a self-affine process whose trajectory has a fractal dimension \cite{falconer,ness}. As seen in the distribution \eqref{fbmdist}, the rescaled FBM process $\gamma^{-\alpha/2}x_\alpha(\gamma t)$ is statistically identical to FBM $\x$ for any scaling factor $\gamma>0$, e.g., $\langle x_\alpha^2(\gamma t) \rangle=\gamma^{\alpha}\la x_\alpha^2(t)\ra$. The fractal dimension of an FBM path can be defined by the walk dimension $d_w$ of the path, i.e., the distance $\sqrt{\vt{x}^2}\sim t^{1/d_w}$ travelled by a random walker. Relating this formula with the scaling behaviour (\ref{msd2}), the walk dimension of FBM $\vt{x}_\alpha(t)$ in $d$-dimensional space is identified as
\begin{equation}
d_w=\frac{2}{\alpha}.\label{dw}
\end{equation}
When $d_w\geq d$, $d$-dimensional FBM is a \emph{compact} (or \emph{recurrent}) random process in the sense that it can visit all sites in the space in the long-time limit. In contrast, when $d_w<d$, $d$-dimensional FBM becomes \emph{non-compact} (or \emph{transient}), and most sites in the space are not visited even in the infinite time limit. In two-dimensional space, subdiffusive FBM with $0<\alpha<1$ is always compact while superdiffusive FBM with $1<\alpha<2$ is non-compact. The difference in the spatial exploration for two-dimensional FBM is depicted in Fig. \ref{fbm1d}.

\section{First passage theory and approximation skills}\label{sec_fpt}

The first passage time is the time $t$ at which a stochastic process crosses
a given threshold value for the first time or escapes from a spatial domain
through a prescribed boundary. The associated first passage time density
$\wp(t)$ is defined via the survival probability $S(t)$ through the relation
\begin{equation}
\wp(t)=-\frac{dS(t)}{dt}.
\end{equation}
The survival probability $S(t)$ is the probability that the random walker survives up to time $t$ within the domain, not being absorbed at the absorbing boundary. In terms of the density $G(\vt{x},t)$ to find the particle at position $\vt{x}$ at time $t$, the survival probability is given by 
\begin{equation}
S(t)=\int_{D}G(\vt{x},t)d^dx.
\label{survpro}
\end{equation}
Because of the absorbing boundary the survival probability decays with time and
$G$ is not normalised, such that $\lim_{t\rightarrow\infty}S(t)=0$ with $S(t=0)
=1$. In principle, once the solution of the density $G$, satisfying the given
boundary conditions, is identified, the survival probability and the first
passage time density can be straightforwardly obtained. However, to actually
perform this calculation is generally nontrivial for many anomalous diffusion
processes \cite{condamin,metzlerctrw,rangarajan,condamin3,aleksei,koren}.

\subsection{Exact result for FBM in one-dimensional domain}

\begin{figure}[tb]
\centerline{\psfig{file=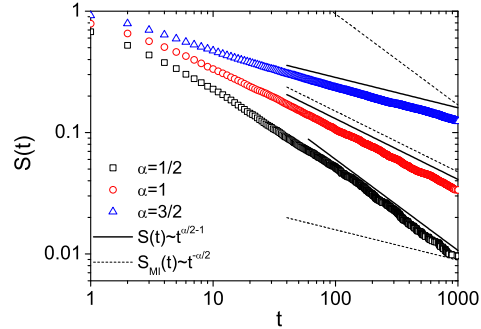,width=7cm}}
\caption{ Numerical results for the survival probabilities $S(t)$ of FBM in a semi-infinite interval with absorbing boundary at $x=0$ for the cases $\alpha=1/2,~1,~3/2$ (from bottom to top). Scaling laws of the survival probability are inserted according to Molchan's result \eqref{spfbm} (solid line) and to the incorrect solution via the method of image (dotted line). } \label{survfbm1D}
\end{figure}

Let us consider the first passage problem of an FBM process in a semi-infinite interval with absorbing boundary at $x=0$. This problem was initially studied by Ding and Yang as a first return problem of an FBM path \cite{ding1}. In this study, as illustrated in Fig.~\ref{fbm1d}(inset), the first return time $t_R$ was defined through the difference of consecutive zero-crossing times $t_R=t_c^{(i+1)}-t_c^{(i)}$ where $t_c^{(i)}$ represents the sequence of the zero-crossing times satisfying $x_\alpha(t_c^{(i)})=0$ of a given FBM path. Using scaling arguments and numerical simulations, it was conjectured that the corresponding first return time density at long times has a power-law decay of the form 
\begin{equation}
\wp(t)\simeq t^{\alpha/2-2}.\label{fptdfbm}
\end{equation}
This conjecture was later confirmed by Molchan's work on the maximum value $M_t$ of FBM on the interval $[0,t]$. It was proven that the probability of $M_t$ being less than a threshold value $c$, say $c=1$, follows the asymptotics 
\begin{equation}
\ln \mathrm{Prob}(M_t<1)=(1-\alpha/2)\ln t^{-1}\left[1+O\Big((\ln t)^{-1/2}\Big)
\right]\label{molchan}
\end{equation}
as $t\rightarrow\infty$ \cite{molchan}. In the context of the first passage process, the probability $\mathrm{Prob}(M_t<c)$ is the survival probability of an FBM not being absorbed up to time $t$ at the location $x=c$ of the absorbing boundary. Therefore, according to \eref{molchan} the long-time scaling law of the survival probability of FBM in semi-infinite interval is identified as \cite{molchan,ding2}
\begin{equation}
S(t)\simeq t^{\alpha/2-1},
\label{spfbm}
\end{equation}
and the corresponding first passage time density is given by the scaling law \eqref{fptdfbm}. Figure \ref{survfbm1D} presents the numerically obtained survival probability $S(t)$ of one-dimensional FBM with anomalous diffusion exponents $\alpha=1/2$ (subdiffusion), 1 (normal diffusion), and $3/2$ (superdiffusion). For all cases Molchan's scaling form \eqref{spfbm} is in good agreement with the simulations results in the long-time limit. 

%%%%%%%%%%%%%%%%%%%%% FIGURE %%%%%%%%%%%%%%%%%%%%%%%%%%%%%%%%%%
%\begin{verbatim}
%%%%%%%%%%%%%%%%%%%%%%%%%%%%%%%%%%%%%%%%%

Let us briefly discuss the first passage behaviour of FBM in a semi-infinite
interval. At $\alpha=1$ it reproduces the universal Sparre-Andersen $3/2$
scaling for symmetric Markovian processes, a special case of which is the
asymptotic behaviour of the L\'{e}vy-Smirnov result for the first passage
density of BM. Somewhat counterintuitively, Molchan's result
suggests that subdiffusive FBM particles with stronger anti-correlation
have more frequent first passage events, while superdiffusive FBM particles
($1<\alpha<2$) tend to survive longer within the confining domain as $\alpha$
gets larger. This is an essential feature of FBM first passage processes,
with profound consequences in concrete applications. An analogous case is the
Kramers escape problem of an FBM particle, studied for an harmonic potential
$U(x)=x^2/2$ with a cutoff (i.e. absorbing boundary) at $x_a\neq0$ in
Ref.~\cite{oleksii}. There, it was numerically shown that the first passage
time density has the exponential
decay $\wp(t)\sim \exp(-t/\la t\ra)$ consistent with the Kramers' theory,
whereas the escape rate $R=\la t \ra^{-1}$ is sensitive to the degree of
persistence or antipersistence of the FBM processes. In accordance to the
first passage behaviour in \eref{fptdfbm} the escape rate $R$ monotonically
increases when FBM is slower ($\alpha$ decreases below unity) while it
decreases as the FBM is faster ($\alpha$ increases above unity). We shall
see below that such first passage behaviour is conserved for FBM processes
in multi-dimensional space.

\subsection{Failure of the method of images}

For Markovian particles, the method of images is a useful technique
to obtain the first passage time density in cases when the geometry of
the problem is relatively simple \cite{redner}. Also it is still
a valid technique for subordinated BMs such as subdiffusive CTRWs
with a diverging mean waiting time \cite{ralfreview2,ralfchapter,metzlerctrw}.
However, it is not applicable to deeply
non-Markovian process of FBM type. To show this, we apply the images method
to the above first passage problem in a semi-infinite domain. Assuming
$G(x,t;x_0)$ is the correct probability density of an FBM particle located
at $x_0>0$ at $t=0$ in a domain $x>0$ with an absorbing boundary at $x=0$,
the method of images yields
\begin{equation}
G(x,t;x_0)=\frac{1}{2\sqrt{4\pi K_\alpha t^\alpha}}\left[\exp \left(-\frac{(x-x_0)^2}{4K_\alpha t^\alpha}\right)-\exp\left(-\frac{(x+x_0)^2}{4K_\alpha t^\alpha}\right)\right].
\end{equation}
Using the survival probability $S(t)=\int_0^\infty G(x,t;x_0)$, it is easily shown that the first passage time density corresponding to this process $G$ has the complete functional form \cite{tobias,jeonwedge}
\begin{equation}
\wp_{\mathrm{MI}}(t)=\frac{\alpha x_0}{\sqrt{4\pi}t^{1+\alpha/2}}\exp\left(-\frac{x_0^2}{4K_\alpha t^\alpha}\right).
\end{equation}
Thus, in the long-time limit $t\rightarrow\infty$, the method of images yields
that the first passage time density scales as
\begin{equation}
\wp_{\mathrm{MI}}(t)\sim t^{-1-\alpha/2}.\label{mi}
\end{equation} 
Except for BM ($\alpha=1$), the scaling of the obtained form does not
agree with the exact result in \eref{fptdfbm}. We demonstrate this fact in
Fig.~\ref{survfbm1D}, in which the survival probability $S_{\mathrm{MI}}(t)\sim
t^{-\alpha/2}$ due to the images method exhibits major deviations in the scaling
behaviour from the  simulation results for both subdiffusive and superdiffusive
cases. The first passage behaviour predicted from $\wp_{\mathrm{MI}}(t)$ is
qualitatively opposite to the correct scaling, in the sense that first
passage events are more frequent as the diffusion is faster ($\alpha$ is
larger). Finally we note that the correct result of Molchan for $\wp(t)$ is
related with the asymptotic scaling of $\wp_{\mathrm{MI}}(t)$ through the
substitution $\alpha\rightarrow2-\alpha$.

\subsection{The pseudo-Markovian approximation} 
The technique of the
pseudo-Markovian approximation\footnote{It is also often referred to
as the Wilemski-Fixman approximation in the literature.}
was originally introduced to describe the cyclisation dynamics of two
end monomers of a polymer chain which is known to be a non-Markovian
process \cite{wf1,wf2,schulten,sokolov2,guerin}. Within the pseudo-Markovian
approximation scheme, the conditional probability $G(\vt{x},t|\vt{x}_0,t_0)$
that the particle is found to be at $\vt{x}$ at time $t$ given that it was
located at $\vt{x}_0$ at time $t_0$ is self-consistently written through
the renewal equation \cite{hughes,redner,sokolov2} \begin{equation}
G(\vt{x},t|\vt{x}_0,0)=\delta(\vt{x}-\vt{x}_0)\delta(t)+\int_0^{t}dt'
\mathcal{P}(\vt{x},t';\vt{x}_0)G(\vt{x},t|\vt{x},t').\label{wfeq}
\end{equation} Here $\mathcal{P}(\vt{x},t;x_0)$ defines the first passage
time density that the diffusing particle embedded in $d$-dimensional space
starts to move at $\vt{x}_0$ at $t=0$ and visits the site $\vt{x}$ for the
first time at time $t'$. For Markovian processes such as BM, Eq.~(\ref{wfeq})
is an exact description relating the conditional probability density $G$
to the corresponding first passage probability \cite{hughes,redner}. In this
case the first passage time density $\mathcal{P}$ in any $d$-dimensional
domain is given by
\begin{equation}
\mathcal{P}(\vt{x},u)=\left\{\begin{array}{ll}
\frac{G(\vt{x},u)}{G(0,u)}, & \vt{x}\neq \vt{x}_0\\ 1-\frac{1}{G(0,u)}, &
\vt{x}=\vt{x}_0\end{array}\right.\label{wflaplace}
\end{equation}
in the
Laplace domain, where $f(\vt{x},u)\equiv \int_0^\infty dt e^{-ut}f(\vt{x},t)$
with $u>0$. For non-Markovian processes, however, it neglects the fact that
the returning probability $G(\vt{x},t|\vt{x},t')$ in the above expression does
depend on all the visited sites of the particle before $t'$. In this
sense, the first passage density $\mathcal{P}$, Eq.~\eqref{wflaplace},
is a Markovian approximation to the true first passage solution for
non-Markovian processes. This technique has been exploited to relate
the first passage problems of FBM and the fractional Langevin equation
associated with FGN \cite{oleksii,chatterjee,bologna,tobias}. It turns
out that the pseudo-Markovian approximation often works reasonably well to
explain the first passage processes for some problem. It is known that the
pseudo-Markovian scheme is the first approximation to the exact perturbative
method introduced by Likhtman and Margues \cite{likthman}.

The first passage problem for FBM in a semi-infinite interval was studied within
the scheme of the pseudo-Markovian approximation \cite{bologna,tobias}. In
particular, in Ref. \cite{tobias} the complete functional form of the
first passage time density was obtained by this method, finding that it
agrees with the simulations over the entire time domain explored. Except
for certain values of $\alpha=(n+1)/(4m+2)$
with $n,m\in \mathbb{Z}^{+}$ the pseudo-Markovian solution of the
corresponding first passage time density, $\wp_{\mathrm{PM}}(t)\equiv
\mathcal{P}(0,t;x_0)$ ($x=0$ is the absorbing boundary), is
found to follow the series expansion \cite{tobias}
\begin{eqnarray}
\wp_{\mathrm{PM}}(t)&=&\frac{\sigma^{-1/\alpha}}{\Gamma(1-\alpha/2)}\left([K(t)]^{2-\alpha}\frac{1}{\alpha}\sum_{n=0}^\infty
[K(t)]^n\frac{(-1)^n}{n!}\frac{\Gamma([\alpha/2-1-n]/\alpha)}{\Gamma(\alpha/2-1-n)}\right.
\nonumber \\ &+& \left. K(t)\sum_{m=1}^\infty[K(t)]^{\alpha
m}\frac{(-1)^m}{m!}\frac{\Gamma(1-\alpha/2-\alpha m)}{\Gamma(-\alpha
m)} \right)
\label{wf}
\end{eqnarray}
with
\begin{equation}
K(t)=\frac{\sigma^{1/\alpha}}{t}.
\end{equation}
Here $\sigma=x_0^2/(4K_\alpha)$ and $\Gamma(z)$ is the Gamma function. In the
long-time limit where $K(t)\rightarrow0$, remarkably, the above pseudo-Markovian
solution \eqref{wf} yields the correct scaling behaviour \eref{fptdfbm} of FBM
\begin{equation}
\wp_{\mathrm{PM}}(t)\sim t^{\alpha/2-2}
\end{equation}
in the
range $1/6\leq\alpha<2$ of the anomalous diffusion exponent. However, it gives
an inconsistent result
\begin{equation}
\wp_{\mathrm{PM}}(t)\sim t^{-\alpha-1}
\end{equation}
for strong anti-persistent FBM with $0<\alpha<1/6$. Numerical
tests showed that $\wp_{\mathrm{PM}}(t)$ describes remarkably well the simulated
first passage behaviour of FBM in the entire time window for the cases
with anomalous diffusion exponent $1/6\leq\alpha\leq1$. However, such
good agreement is not fulfilled for superdiffusive FBM with $1<\alpha<2$,
where $\wp_{\mathrm{PM}}(t)$ only correctly captures the long-time scaling
law \eqref{fptdfbm}.

\subsection{First passage process of fractional Langevin equation}

%%%%%%%%%%%%%%%%%%%%% FIGURE %%%%%%%%%%%%%%%%%%%%%%%%%%%%%%%%%%
%\begin{verbatim}
\begin{figure}[tb]
\centerline{\psfig{file=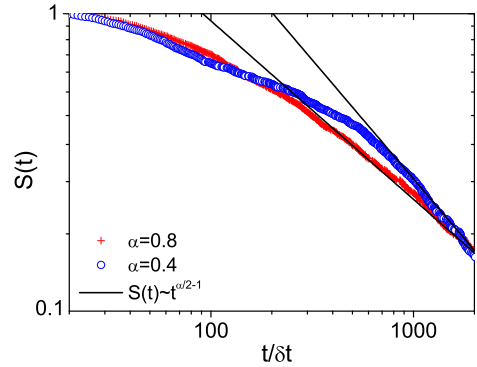,width=7cm}}
\caption{Numerical results for the survival probabilities $S(t)$ of FLE motion with anomalous diffusion exponent $\alpha=0.4$ and 0.8 (from top to bottom) on a semi-infinite interval with absorbing boundary at $x=0$. Here the anomalous diffusion exponent $\alpha$ is associated with the FLE motion $y_\alpha(t)$ satisfying the scaling relation $\la y_\alpha^2(t)\ra\sim t^\alpha$ in the overdamped limit, see Sec.~\ref{sec_fbm}. The predicted scaling law $S(t)\sim t^{\alpha/2-1}=t^{-\alpha'/2}$ compares favourably with the simulation results. Parameter values in the simulation: integration time step $\delta t=0.001$, particle mass $m=0.1$, generalised frictional coefficient $\gamma=100$, and $k_B\mathcal{T}=1$.} \label{spfle1d}
\end{figure}
%%%%%%%%%%%%%%%%%%%%%%%%%%%%%%%%%%%%%%%%%%

As mentioned previously, the overdamped FLE process $y_\alpha(t)$ driven by a persistent FGN $\xi_{\alpha'}(t)$ with $1<\alpha'<2$ behaves analogously to a subdiffusive FBM $\x$ of $0<\alpha<1$ provided that $\alpha=2-\alpha'$ \cite{deng,taloni}. Both processes share the same stochastic properties, and thus one can identify the first passage behaviour of FLE motion from that of its counterpart FBM process, and vice versa \cite{taloni}. Using this correspondence, we find that the survival probability and the associated first passage time density of the overdamped FLE process $y_\alpha(t)$ in a semi-infinite interval are respectively governed by Molchan's scaling results \eqref{spfbm} and \eqref{fptdfbm}, i.e., the first passage process of FLE motion is governed by the asymptotic laws
\begin{subequations}
\label{fptfle}
\begin{eqnarray}
S(t)&\simeq& t^{-\alpha'/2},\\
\wp(t)&\simeq& t^{-\alpha'/2-1}
\end{eqnarray}
\end{subequations}
in terms of the anomalous diffusion exponent $\alpha'$ of the persistent
FGN. We emphasise here that the above scaling forms \eqref{fptfle} should
not be confused with \eref{mi} obtained by the method of images, albeit they
appear to have the identical exponents. In other words, the scaling form
\eqref{fptfle} does not imply that the method of images can be successfully
applied to FLE motion.

The first passage time behaviour \eref{fptfle} of the FLE process was
numerically validated in terms of the single-file diffusion scheme, which is
known to be
governed by FLE motion $y_\alpha(t)$ of the anomalous diffusion exponent
$\alpha=1/2$ \cite{taloni,tobias}. For further numerical confirmation of
the equivalence of the first passage behaviour between FBM and FLE motion,
in Fig.~\ref{spfle1d} we simulate the first passage process of FLE motion
$y_\alpha(t)$ with $\alpha=0.4$ and 0.8. It shows that the long-time scaling
behaviour of FLE processes are in good agreement with Molchan's scaling
laws [i.e., \eref{fptfle}]. It is worthwhile mentioning that on short time
scales, when inertia effects are relevant, the first passage behaviour of
FLE motion naturally deviates significantly from those of FBM, as shown in
our simulations.

\begin{figure}[tb]
\centering
\psfig{file=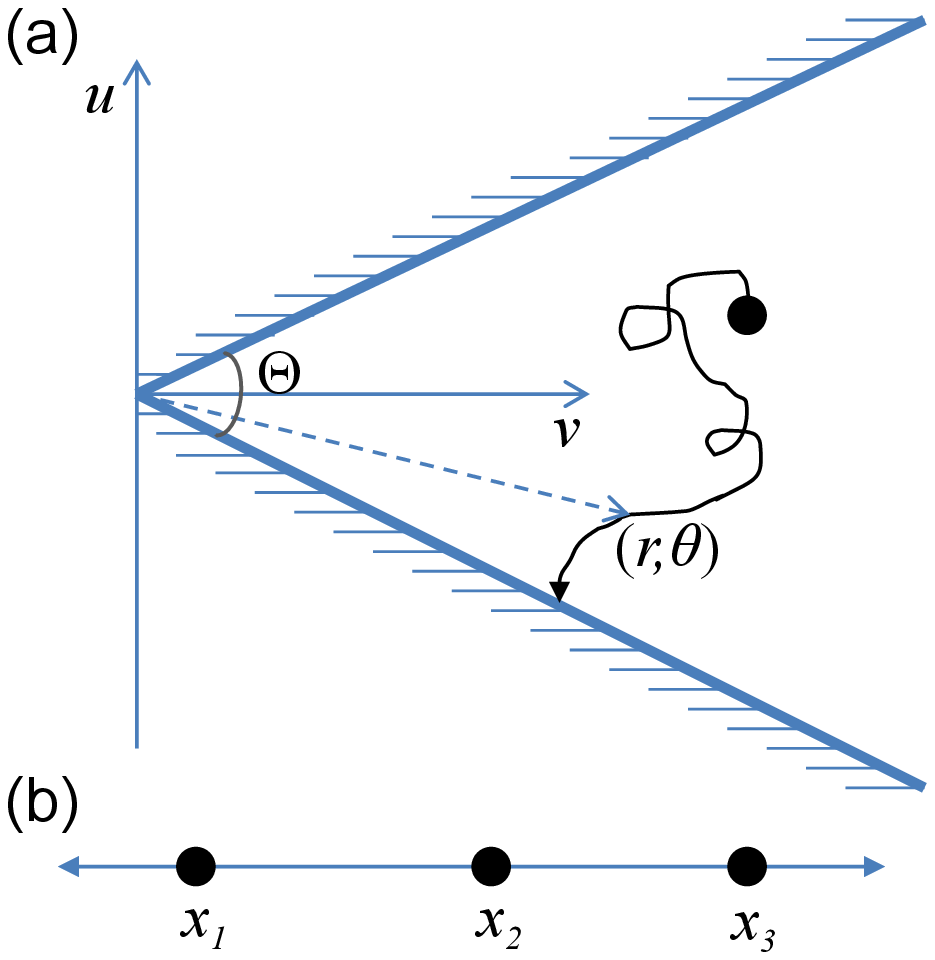,width=4.5cm}
\psfig{file=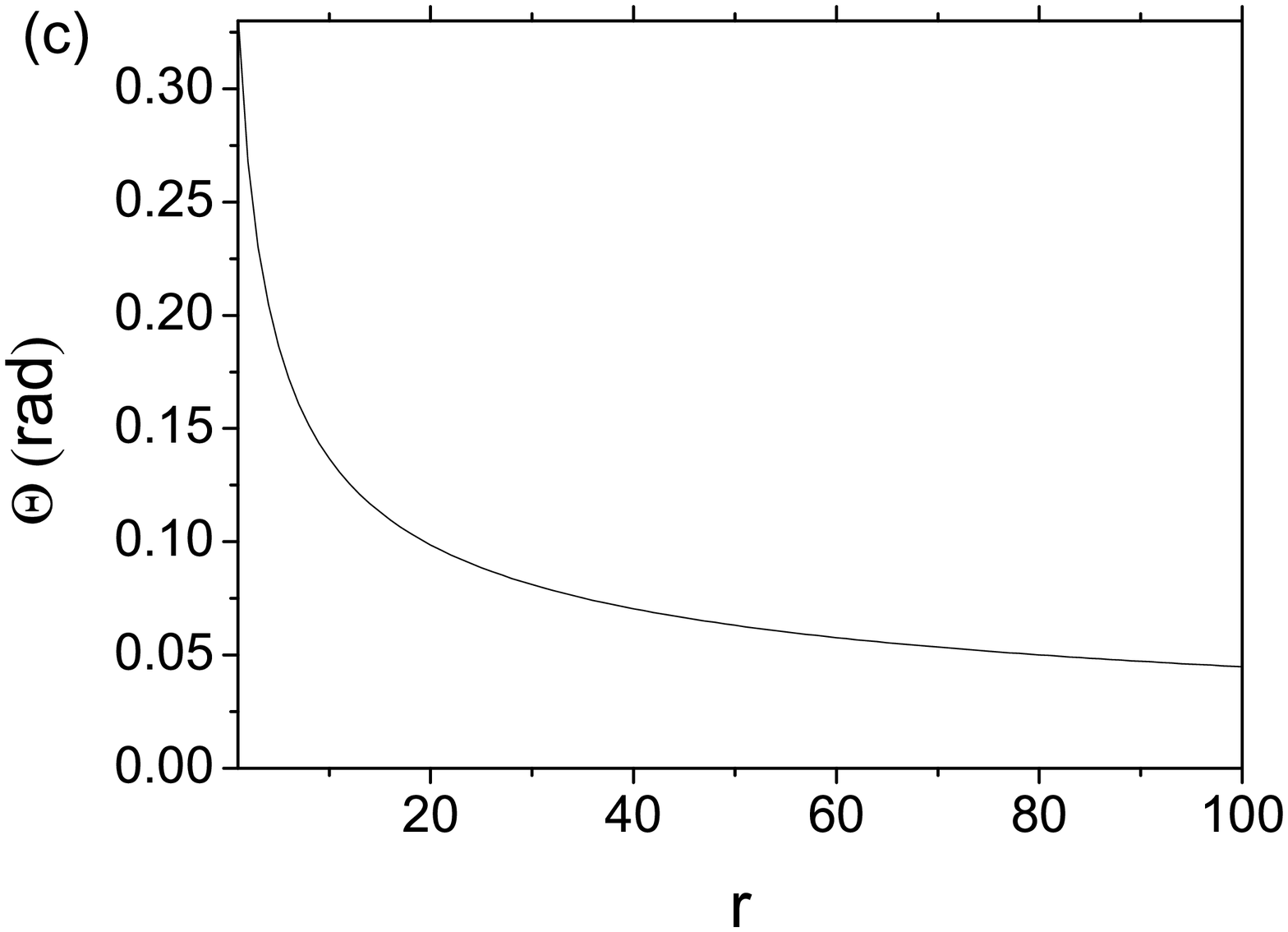,width=5.5cm}
\caption{(a) First passage process of
an FBM particle confined to a two-dimensional open wedge of angle $\Theta$
with absorbing walls. The trajectory of the particle is described in
polar coordinates by its radius $r$ and angle $\theta$.
(b) Three-body reaction of FBM particles in a one-dimensional
interval. The location of the $i$th particle is given by $x_i$. (c)
Dependence of the wedge angle $\Theta$ as a function of the ratio
$r=K^{(2)}_\alpha/K^{(1)}_\alpha$ in (b). See text for details.}
\label{figwedge}
\end{figure}

\section{Diffusion-limited reaction of FBM particles}\label{application}

Based on the first passage behaviour of an FBM particle studied in the previous section, here we explore the kinetics of diffusion-limited reaction for FBM particles in $d$-dimensional space. Under the assumption that a reaction spontaneously occurs whenever two diffusing reactants meet, our reaction problem can be regarded as the problem of a stochastic searcher to find one of multiple target sites distributed in space.

\subsection{Three-body reactions in one-dimensional space}

Before we embark for the multi-particle case, let us consider the first passage
problem of a single FBM particle that is confined to a two-dimensional open
wedge of angle $\Theta$ [see Fig.~\ref{figwedge}(a)]. The particle is assumed
to escape out of the wedge when it hits its absorbing wall. It turns out that
this wedge problem is mapped onto the diffusion-limited three-body reaction 
problem in
one-dimensional space \cite{redner,avraham,redner2,jeonwedge}. In the special
case of BM with $\alpha=1$, the first passage statistics is analytically
solvable using the Green's function formalism. For the general case of FBM
with $\alpha\neq1$ the first passage statistics is not analytically derivable
from an equation, but it can be conjectured from combination of Molchan's result
for the
one-dimensional problem and the exact solution at $\alpha=1$ \cite{jeonwedge}.

\subsubsection{Case I: Exact solution for BM \label{secwedgeapprox1}}

The motion of a diffusing particle in an absorbing wedge [Fig.~\ref{figwedge}(a)]
satisfies the diffusion equation in polar coordinates,
\begin{equation}
\frac{\partial}{\partial t}G(r,\theta;t)=K_{1}\left(\frac{\partial^2}{\partial
r^2}+\frac{1}{r}\frac{\partial}{\partial r}+\frac{1}{r^2}\frac{\partial^2}{
\partial\theta^2}\right)G(r,\theta;t)
\end{equation}
with the boundary conditions $G(r,0;t)=G(r,\Theta;t)=0$ with $0<\Theta<2\pi$ and the initial condition $G(r,\theta;0)=\delta(r-r_0)\delta(\theta-\theta_0)/r_0$. It can be shown that at large time $t$ the solution $G$ has the asymptotic form \cite{redner}
\begin{equation}
G(r,\theta;t)\simeq\frac{\pi\sin\left(\frac{\pi\theta}{\Theta}\right)}{4K_{1}
\Theta t}\exp\left(\frac{-r^2+r_0^2}{4K_{1}t}\right)I_{\pi/\Theta}
\left(\frac{rr_0}{2K_{1}t}\right),
\label{Gsolution}
\end{equation}
where $I_\nu$(z) is the modified Bessel function of the first kind. Using the solution \eqref{Gsolution} we are able to obtain the survival probability of the Brownian particle $S(t)=\int_0^\Theta\int_0^\infty rG(r,\theta;t)drd\theta$ and the corresponding first passage time density $\wp(t)=-\frac{dS(t)}{dt}$. From the approximations $I_\nu(z)\approx(z/2)^\nu/\Gamma(1+\nu)$ and $e^{-r_0^2/[4K_{1}t]}\approx 1$ we further obtain the long-time scaling forms of the two quantities such that
\begin{subequations}
\label{fptdH05}
\begin{eqnarray}
S_{\Theta}(t)&\simeq&\left(\frac{r_0}{\sqrt{K_{1}}}\right)^{\pi/\Theta}t^{-\pi/(2
\Theta)},\\[0.2cm]
\wp_\Theta(t)&\simeq&\frac{\pi}{2\Theta}\left(\frac{r_0}{\sqrt{K_{1}}}\right)^{\pi/\Theta}
t^{-1-\pi/(2\Theta)}.
\end{eqnarray}
%\endnumparts
\end{subequations}
The obtained expressions reproduce the L\'{e}vy-Smirnov scaling form
for BM in a semi-infinite interval when the wedge becomes half-infinite
at $\Theta=\pi$, as it should be. It also gives an important general
relation of the survival probability for the wedge of a right angle,
$S_{\pi/2}(t)=S_{\pi}^2(t)$, expected from the independence of $x$ and
$y$ motion. Importantly, the scaling exponents depend on the wedge angle
$\Theta$ inverse-proportionally. The first passage behaviour in a wedge of
angle $\Theta<\pi/2$ is qualitatively different from those in a wedge of
angle $\Theta\geq\pi/2$; in the former case the mean first passage time
\begin{equation}
\la t \ra=\int_0^\infty t\wp_\Theta(t)dt \label{mfpt}
\end{equation}  
is finite while it diverges for the latter case.

\subsubsection{Case II: Exact solution for FBM with wedge angles $\Theta=\pi$
and $\Theta=\pi/2$\label{secwedgeapprox2}}

From Molchan's exact results \eqref{fptdfbm} and \eqref{spfbm}, the scaling
of the first passage of FBM confined to a wedge of angle $\pi$ due to the
independence of spatial directions is necessarily
\begin{subequations}
\begin{eqnarray}
S_{\pi}(t)&\simeq& t^{\alpha/2-1},\\
\wp_{\pi}(t)&\simeq& t^{\alpha/2-2}
\end{eqnarray}
\end{subequations} 
for any exponents $0<\alpha<2$. Making use of the general relation of the survival probability $S_{\pi/2}(t)=S_{\pi}^2(t)$, we also find the exact scaling forms of the first passage events for a wedge of angle $\Theta=\pi/2$,
\begin{subequations}
\label{fptd90}
\begin{eqnarray}
S_{\pi/2}(t)&\simeq& t^{\alpha-2},\\
\wp_{\pi/2}(t)&\simeq& t^{\alpha-3}.
\end{eqnarray}
\end{subequations}

\subsubsection{Case III: Conjecture for FBM with wedge angle $0<\Theta<2\pi$}

Using the exact first passage scaling for the special cases presented in
Secs.~\ref{secwedgeapprox1}\& \ref{secwedgeapprox2} we now conjecture
its complete functional form for FBM confined to a wedge of any angle
$0<\Theta<2\pi$. To get a clue for its scaling form, we start
to derive the survival probability and the first passage time density for
the scaled Brownian motion process
$\overline{\mathbf{x}}_\alpha(t)$ whose probability
density $\overline{G}(\mathbf{x},t;\vt{x}_0)$ satisfies the generalised
diffusion equation \cite{luts}
\begin{equation}
\frac{\partial}{\partial t}\overline{G}(\mathbf{x},t;\vt{x}_0)=K(t)\nabla^2\overline{G}(\mathbf{x},t;\vt{x}_0).
\label{diffusion2}
\end{equation} 
Here the time-dependent diffusion constant is given by $K(t)\sim K_1 t^{\alpha-1}$ so that the mean squared displacement of the process has the scaling behaviour $\langle\overline{\mathbf{x}}_\alpha^2(t)\rangle\simeq 4K_1 t^\alpha$. It is known that the scaled Brownian motion process described by \eref{diffusion2} is non-stationary \cite{sokolov}, thus not an FBM. Following the steps introduced in Sec. \ref{secwedgeapprox1} we find that the first passage time quantities of this process are
\begin{subequations}
\label{wrong}
\begin{eqnarray}
\overline{S}_\Theta(t)&\propto& t^{-\pi \alpha/(2\Theta)},\\
\overline{\wp}_\Theta(t)&\propto& t^{-1-\pi \alpha/(2\Theta)}.\label{wrongfptd}
\end{eqnarray}
\end{subequations}
The obtained scaling behaviours are identical to those for BM when $\alpha=1$, but inconsistent with the results of FBM for wedges of $\Theta=\pi$ and $\Theta=\pi/2$. Albeit incorrect, from \eqref{wrong} we infer that the correct scaling for FBM is also expressed in terms of the product of $\alpha$ and $\Theta^{-1}$. Under this assumption, there exists only one unique functional form that is consistent with all of the above special cases, and this is our conjecture for the first passage time quantities of FBM confined to a 2D wedge: 
\begin{subequations}
\begin{eqnarray}
S_\Theta(t)&\simeq& t^{\pi(\alpha-2)/(2\Theta)},\label{S}\\
\wp_\Theta(t)&\simeq& t^{-1+\pi(\alpha-2)/(2\Theta)}.
\end{eqnarray}
\label{fptd}
\end{subequations}

The validity of this conjecture was numerically tested in Ref.~\cite{jeonwedge}. We note that through the correspondence $\alpha\rightarrow2-\alpha$ the conjecture \eqref{fptd} is obtained from \eref{wrong}, in the same way as between the solution \eqref{fptdfbm} by Molchan and the approximation \eqref{mi} by the images method for the one-dimensional case. Also consistent to the first passage behaviour discussed in the previous section, the conjecture \eqref{fptd} states that first passage events occur more frequently as the FBM has stronger anti-persistent memory (smaller $\alpha$). The mean first passage time $\la t \ra$ defined by \eref{mfpt} is finite for wedge angles $\Theta<\Theta_c$ where the critical angle 
\begin{equation}
\Theta_c=\pi\left(1-\frac{\alpha}{2}\right)=\pi\left(\frac{d_w-1}{d_w}\right)
\end{equation} 
depends on $\alpha$, or, alternatively on the walk dimension $d_w$ of the FBM.

\subsubsection{Reaction kinetics of three FBM particles}

Consider a coalescence reaction of $N$ FBM particles diffusing in
one-dimensional space. For simplicity, it is assumed that a reaction occurs
whenever any two of them meet, and after the reaction the coalesced
particle (or cluster) is also governed by FBM. As for its reaction dynamics,
this problem can be reduced to a three-body problem where the middle FBM
particle is surrounded by two coalesced FBM particles. The survival condition
of the middle particle is given by two inequalities, $x_1<x_2$
and $x_2<x_3$, where $x_i$ represents the position of the $i$th particle [see
Fig.~\ref{figwedge}(b)]. To solve the problem we assume that $\left(\frac{x_1(t)}{
\sqrt{K^{(1)}_\alpha}},\frac{x_2(t)}{\sqrt{K^{(2)}_\alpha}},\frac{x_3(t)}{\sqrt{
K^{(3)}_\alpha}}\right)$ is the
rescaled Cartesian component of a single FBM particle $\vt{x}_\alpha(t)$ in a
three-dimensional space, where $K^{(i)}_\alpha$ defines the generalised
diffusivity of the $i$th particle in the above one-dimensional space
\cite{redner,avraham}. Then, the original survival problem of the three
one-dimensional particles can be understood such that the \emph{single\/}
three-dimensional FBM particle diffuses inside the domain defined by the
intersects of two absorbing planes $x_1=x_2$ and $x_2=x_3$, in general with
different diffusivities in the different spatial directions \cite{redner}. Since
the particle's motion in the direction perpendicular to both normal vectors of
the two planes is independent of the absorbing event, the survival problem in
the three-dimensional domain reduces to the two-dimensional wedge problem
studied above. Here the wedge angle in the new orthonormal coordinate $(u,v)$
on the plane spanned by the two normal vectors can be unambiguously found to be
\begin{equation}
\Theta=\arccos\left(\frac{K^{(2)}_\alpha}{\sqrt{\left(K^{(1)}_\alpha+K^{(2)}_
\alpha)(K^{(2)}_\alpha+K^{(3)}_\alpha\right)}}\right)
\end{equation}
from the geometry of the confining domain \cite{redner}.

Let us discuss the meaning of these results: (1) In the case when all three particles have identical diffusivity $K^{(1)}_\alpha=K^{(2)}_\alpha=K^{(3)}_\alpha$, the wedge angle is given by $\Theta=\pi/3$. Then the survival probability of the middle particle decays as
\begin{equation}
S(t)\simeq t^{-3(2-\alpha)/2}.
\end{equation}   
It reproduces the classical result $S(t)\sim t^{-3/2}$ of the decay of the concentration of the remaining reactant particles for $\alpha=1$ \cite{avraham}. The decay law is modified for FBM particles in a seemingly counterintuitive fashion: for the superdiffusive FBM particle the reaction occurs more slowly, resulting in a slower decay of the surviving particles. In contrast, the subdiffusive FBM particles have increased likelihood to meet each other and produce a sharp decay of the survival probability. Indeed, this property is related to the higher recurrence
probability of an FBM motion discussed above.

(2) In the case when the middle particle is immobile $K^{(2)}_\alpha=0$, the wedge angle is $\Theta=\pi/2$. It corresponds to the situation that the two surrounding particles independently search their target and thus the decay law of the survival probability is given by
\begin{equation}
S(t)=S_{\pi}^2(t)\simeq t^{-(2-\alpha)}.
\end{equation}
For both subdiffusive and superdiffusive motions the general tendency is that
the decay of the survival probability is always slowed down by the immobilisation
of the target particle. In other words, a moving target always has a higher
likelihood
to be captured.

(3) The surrounding particles have an identical diffusivity
$K^{(1)}_\alpha=K^{(3)}_\alpha$ and their diffusivity is smaller than that of
the middle particle, $K^{(1)}_\alpha<K^{(2)}_\alpha$. A physical scenario of
this case is that the surrounding particles are already coalesced FBM particles
with a decreased diffusivity due to the larger particle size.
The dependence of the wedge angle
$\Theta=\cos^{-1}[r/(1+r)]$ on the ratio $r=K^{(2)}_\alpha/K^{(1)}_\alpha$
is plotted in Fig.~\ref{figwedge}(c). The fact that $\Theta$ monotonically
decreases from $\pi/3$ to $0^+$ as $r$ increases from 1 to infinity implies that
slower movement of the surrounding particles always leads to an increased
hitting probability between the middle and surrounding particles. In a
naive expectation it may be assumed that the diffusivity of the surrounding
particles is decreased inverse-proportionally to the number of the coalesced
particles. Then, the profile of $\Theta$ in Fig.~\ref{figwedge}(c) can be
understood such that their reaction kinetics is noticeably  accelerated only
up to a few tens of particle coalescence events.

\subsection{Many-body reactions of multi-dimensional FBM particles}

In the previous section we studied the reaction kinetics of one-dimensional
FBM particles using the analogy to  the first passage process of an FBM particle that escapes from a confined $d$-dimensional domain through absorbing walls. For
the study of the reaction kinetics of $d$-dimensional FBM (with $d\geq2$),
we now consider the first arrival problem in free $d$-dimensional space that
an FBM particle with the initial condition $\vt{x}_\alpha(0)=0$ arrives at a
specific target site $\vt{x}=\vt{x}_T$ at time $t$ for the first time. For
BM this problem is exactly solvable using the renewal equation \eqref{wfeq}
\cite{redner}. For non-Markovian FBM, to our knowledge,
this question represents an open problem. Here, we employ the pseudo-Markovian  approximation
scheme to find the scaling laws of the corresponding first arrival time
quantities for FBM with $d>1$ and test their validity to simulation results.

To do this, let us first consider the Laplace transform of $G(\vt{x}=0,t)=(4\pi K_\alpha t^\alpha)^{-d/2}$ of a free $d$-dimensional FBM. Using the relation \eqref{dw} between the anomalous diffusion exponent $\alpha$ and the walk dimension $d_w$ and $z\equiv \exp(-u)$ we arrive at
\begin{equation}
G(0,z)=\int^\infty_{l_c} G(\vt{x}=0,t)e^{-ut}dt=(4\pi K_\alpha)^{-d/2}\int^\infty_{l_c} dt t^{-\frac{d}{d_w}}z^t.\label{G0z}
\end{equation}
Here the lower cutoff $l_c$ of the order of unity is introduced to prevent the physically irrelevant divergence of the integral as $t\rightarrow0$, which does not affect the long-time behaviour. This integral behaves differently depending on the ratio $d/d_w$: (i) For $d<d_w$ (compact exploration), the integral diverges like $G(0,z)\sim A_d(1-z)^{(d-d_w)/d_w}$ with a proportionality factor $A_d=(\frac{d}{d_w-d})(4\pi K_\alpha)^{-d/2}$ as $z\rightarrow1$ (equivalently $u\rightarrow0$). (ii) For $d=d_w$, \eref{G0z} diverges logarithmically $G(0,z)\sim -A_{dw}\log(1-z)$ where $A_{dw}=(4\pi K_\alpha)^{-d_w/2}$. (iii) For $d>d_w$ (non-compact exploration), $G(0,z)$ converges to $(1-\mathcal{R})^{-1}+B_d(1-z)^{(d-d_w)/d_w}$, where $\mathcal{R}$ turns out to be the non-vanishing probability that the random walk eventually returns to the starting site $\vt{x}=0$ and $B_d=d_w/(d-d_w)$. Following the general strategy presented in Ref.~\cite{redner} of finding the long-time scaling behaviour of first passage processes from $G(0,z)$, it is straightforward to show that the survival probability behaves as
\begin{eqnarray}
S(t)\sim \left\{\begin{array}{ll} A_d^{-1}t^{\frac{d-d_w}{d_w}}, & d<d_w\\[0.2cm]
A_{dw}^{-1}(\log t)^{-1}, & d=d_w \\[0.2cm]
(1-\mathcal{R})+C_d t^{\frac{d_w-d}{d_w}}, & d>d_w
\end{array}\right.\label{stfat}
\end{eqnarray}
and the corresponding first arrival time density becomes
\begin{eqnarray}
\wp(t)\sim \left\{\begin{array}{ll} t^{\frac{d}{d_w}-2}, & d<d_w\\[0.2cm]
t^{-1} (\log^2 t)^{-1}, & d=d_w \\[0.2cm]
t^{-\frac{d}{d_w}}, & d>d_w
\end{array}\right. .\label{fatd}
\end{eqnarray}
For the case of the compact exploration $d<d_w$ and the borderline case
$d=d_w$, the survival probability eventually decays to zero, but the mean
first arrival time $\la t\ra=\int^\infty_{l_c} tp(t)dt$ is always infinite
irrespective of the ratio $d/d_w\leq 1$. It reflects the fact that the
diffusing particle always finds its target only in the limit of infinite
time. Consequently, in the reaction kinetics scenario the compact exploration of
a reactant particle facilitates local reaction with the nearby reactant
particles while it eventually slows down the total reaction kinetics over the
space by preventing the well-mixing of the reactant particles. Consistent
results were reported in the numerical study of reaction kinetics of FBM
particles in two-dimensional space \cite{hellmann}. Note that the slow-down
of the reaction kinetics of the compact-exploring FBM particle is mainly due
to the effect of higher dimensionality $d\geq 2$. When $d=1$, reactions only
occur among neighbouring particles and thus the reaction kinetics becomes
faster for compact-exploring FBM particles. This is also consistent with
the fact that they have a finite mean first passage time.

For non-compact exploring FBM particles with $d>d_w$, the survival
probability decays towards a nonzero constant $1-\mathcal{R}$ as
$t\rightarrow\infty$. Thus, the non-compact searcher never locates its target
with a probability $1-\mathcal{R}$, even in the limit of $t\rightarrow\infty$.
For the particles contributing to the reaction with a probability $\mathcal{R}$,
the mean first arrival time $\la t \ra$ becomes finite if $d_w<d/2$. In
this case, the non-compact exploration enables a reactant particle to find
distant counterpart particles.

\begin{figure}[tb]
\centering
\psfig{file=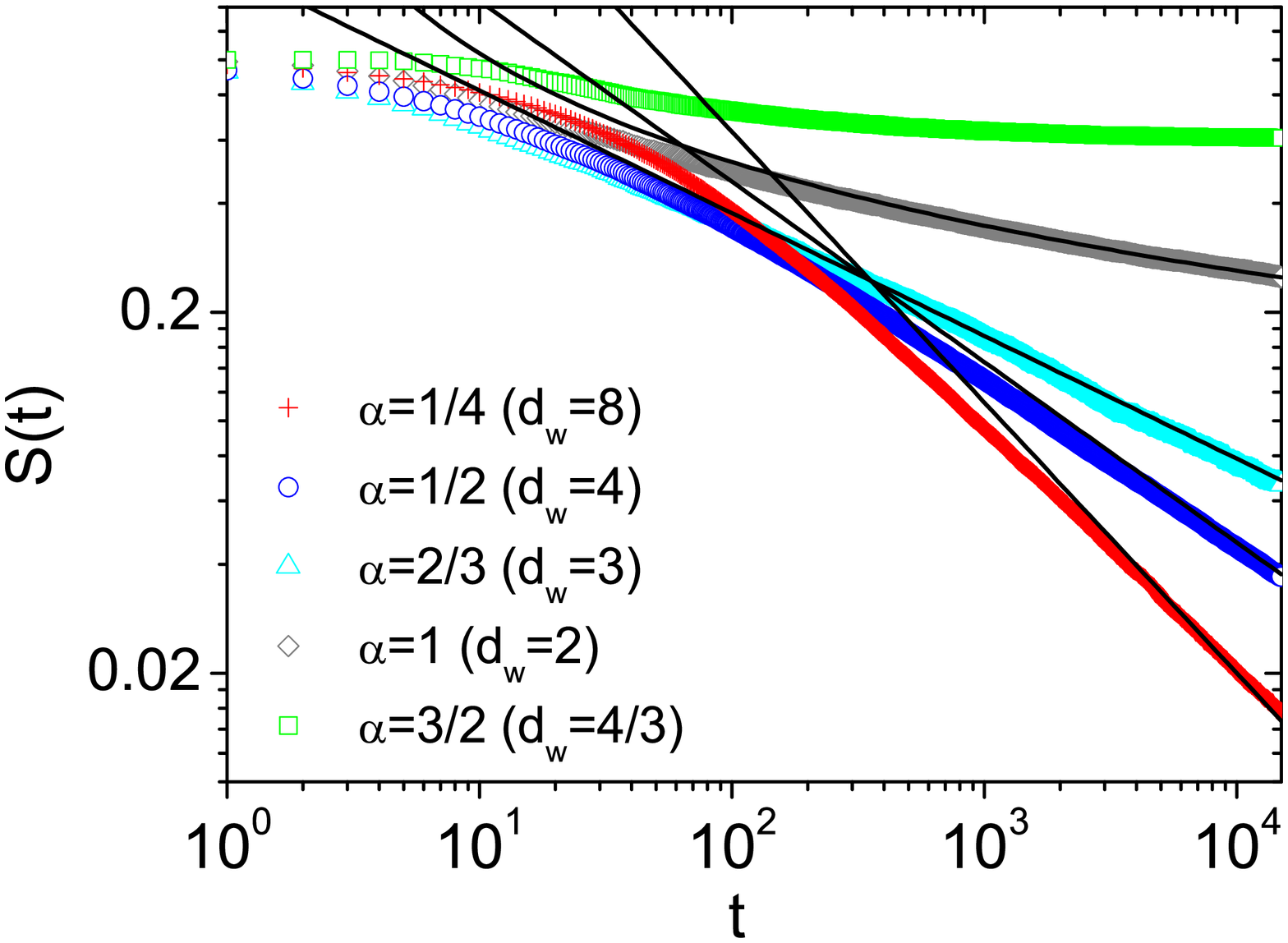,width=5.3cm}
\psfig{file=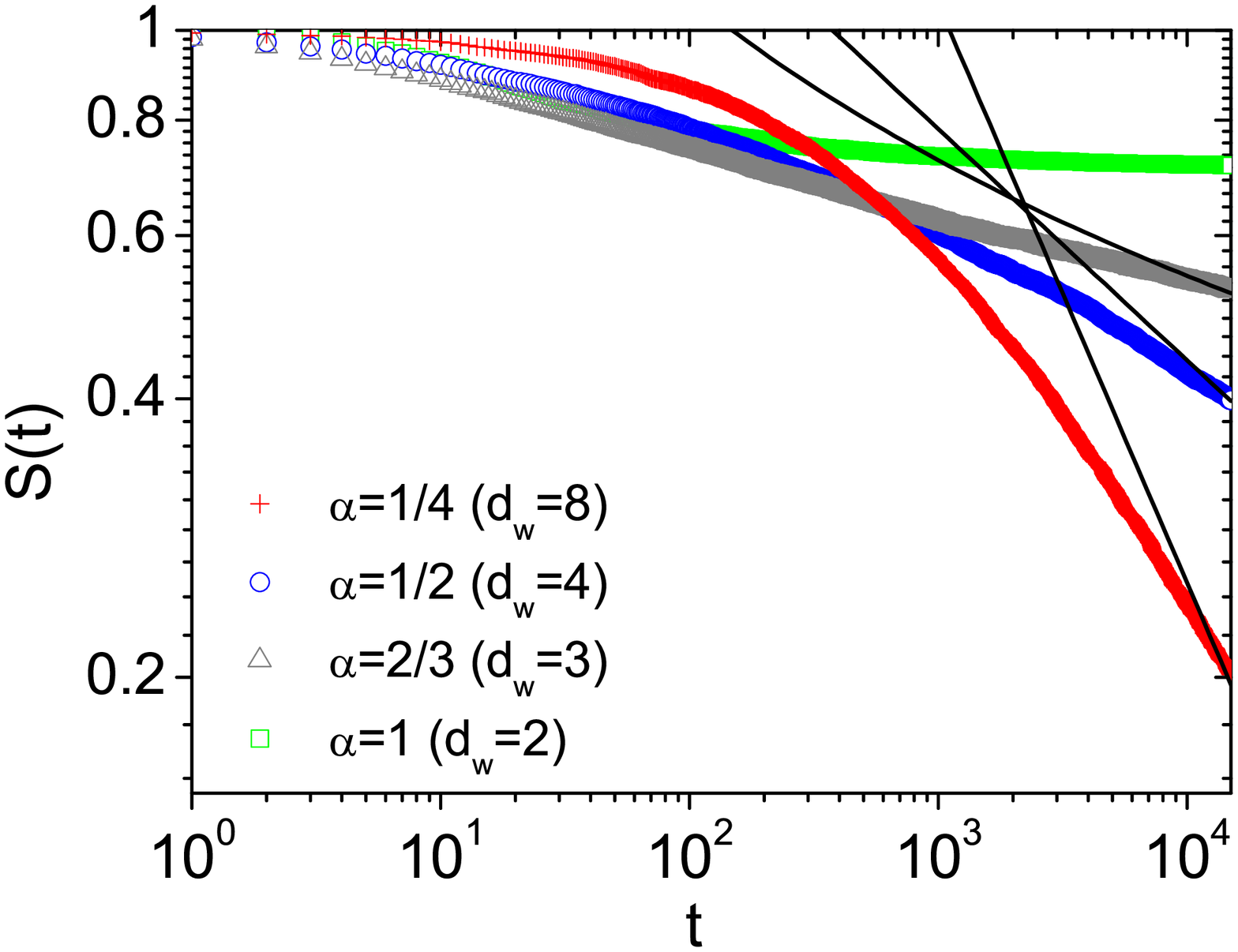,width=5.3cm}
\caption{Survival
probability $S(t)$ of a $d$-dimensional FBM starting at $\vt{x}=0$ and not being
absorbed at $\vt{x}=\vt{x}_T(\neq 0)$ up to time $t$, (Left) for $d=2$ and
(Right) for $d=3$. The first arrival processes of FBM with different values
of $\alpha$ are simulated to check all cases: $d<d_w$ (compact exploration),
$d=d_w$, and $d>d_w$ (non-compact exploration). Left: two-dimensional FBM with
$\alpha=1/4,~1/2,~2/3,~1$, and 3/2 (bottom to top). Right: three-dimensional
FBM with $\alpha=1/4,~1/2,~2/3$, and 1 (bottom to top). In each panel, solid
lines depict the solution using the pseudo-Markovian approach \eqref{stfat}.}
\label{fat2d3d}
\end{figure}

For BM with walk dimension $d_w=2$, the above pseudo-Markovian results
\eqref{stfat} and \eqref{fatd} are evidently exact in any spatial dimension
\cite{redner}. In the case of
FBM in a one-dimensional space, the pseudo-Markovian approximations
are also consistent with Molchan's first passage time results. Are the
approximation still sufficiently accurate in explaining at least the long-time
scaling laws of the first arrival processes for $d$-dimensional FBM with
$d\geq2$? We numerically investigate this with simulations of the first arrival
processes of FBM in two- and three-dimensional space. Figure \ref{fat2d3d}
presents the comparison of the corresponding survival probabilities
between simulations and the pseudo-Markovian scaling law \eqref{stfat}. 
A notable general
tendency is that the pseudo-Markovian approximation gets worse as the
dimensionality is increased. In $d=3$ the power-law exponents of $S(t)$
are significantly different from $(d-d_w)/d_w$ for compact exploration
and $S(t)$ does not logarithmically decay for $d_w=d(=3)$. In $d=2$
the pseudo-Markovian approximation produces reasonable results compared to
the simulations. Moreover,
the theory tends to be exact as $d_w\rightarrow d$, at which $S(t)$ decays
logarithmically with time $t$.

The fractal kinetics of a reaction $A+B\rightarrow C$ for subdiffusive
FBM reactants in a two-dimensional space was numerically studied in
Ref.~\cite{hellmann}. The reaction kinetics was shown to follow the
general rule $[C]=k_0 t^{-h}[A][B]$ with $0\leq h\leq1$ and a constant
$k_0$. The parameter $h$ tends to increase as the FBM reactant has a stronger
anti-persistency ($\alpha$ is decreased). The simulation also showed that
the subdiffusive character of the reactants leads to strong segregation of
reactants and, in turn, slower sampling process. As a consequence, the total
amount of products, $N_C(t)=N(0)-N_A(t)-N_B(t)$, decreases slower as $\alpha$
is decreased. These results are in line with the reaction kinetics obtained
from the approximation above.
Clearly, a full understanding of this topic requires further
studies.

\section{Conclusions}\label{conclusion}

In this chapter we provided an overview of some recent progress in the
theoretical understanding of the first passage time behaviour of FBM in one-
and higher-dimensional space. We also introduced an application to
the kinetics of diffusion-limited reaction of FBM reactants. We showed that
the first passage processes of FBM (and overdamped FLE motion)
qualitatively possess generic features in the sense that the first passage
events tend to occur more frequently when FBM attains stronger antipersistence
in its displacement correlation, in other words, when the anomalous diffusion
exponent $\alpha$ of the motion decreases. Our simulations suggest
that the same tendency still holds for multi-dimensional FBM, but in spaces
of dimension higher than one the first passage processes turn out to possess
three distinct statistical regimes depending on the ratio of the dimensionality
of the embedding space and the walk dimension associated with FBM. We found
that the first passage behaviour is qualitatively explained in terms of the
pseudo-Markovian
approximation of the first passage processes. However, the scaling law of
the first passage time quantities are quantitatively consistent with the
pseudo-Markovian approximation only for the one-dimensional case. Intriguingly,
the pseudo-Markovian results tend to  deviate more from the simulations
results as the dimensionality of the space is increased, albeit each component
of a multi-dimensional FBM is independent. This discrepancy is worthwhile
investigating in detail in the future.

\section*{Acknowledgements}
This work is supported by the Academy of Finland within the Finland
Distinguished Professor (FiDiPro) programme. We gratefully acknowledge
fruitful discussion with Otto Pulkkinen and Michael Lomholt. 

\bibliographystyle{ws-rv-van}
%\bibliography{ws-rv-sample}

%\printindex[aindx]                 % to print author index
\printindex                         % to print subject index
\end{document}